\documentclass[aps,showpacs,twocolumn,prl,superscriptaddress]{revtex4}

\usepackage{amsmath}
\usepackage{latexsym}
\usepackage{graphicx}
\usepackage{verbatim}

\begin{document}

\title{Consistency of post-Newtonian waveforms with numerical relativity}

\author{John G. Baker}
\affiliation{Gravitational Astrophysics Laboratory, NASA Goddard Space Flight Center, 8800 Greenbelt Rd., Greenbelt, MD 20771, USA}
\author{James R. van Meter}
\affiliation{Gravitational Astrophysics Laboratory, NASA Goddard Space Flight Center, 8800 Greenbelt Rd., Greenbelt, MD 20771, USA}
\affiliation{Center for Space Science \& Technology, 
University of Maryland Baltimore
County, Physics Department, 1000 Hilltop Circle, Baltimore, MD 21250}
\author{Sean T. McWilliams}
\affiliation{University of Maryland, Department of Physics, College Park, MD 20742, USA}
\author{Joan Centrella}
\author{Bernard J. Kelly}
\affiliation{Gravitational Astrophysics Laboratory, NASA Goddard Space Flight Center, 8800 Greenbelt Rd., Greenbelt, MD 20771, USA}

\date{\today}

\begin{abstract}
General relativity predicts the gravitational wave signatures of
coalescing binary black holes. Explicit waveform predictions for such
systems, required for optimal analysis of observational data, have so
far been achieved using the post-Newtonian (PN) approximation.  The
quality of this treatment is unclear, however, for the important
late-inspiral portion.  We derive late-inspiral waveforms via a
complementary approach, direct numerical simulation of Einstein's
equations. We compare waveform phasing from simulations of the last
$\sim 14$ cycles of gravitational radiation from equal-mass,
nonspinning black holes with the corresponding 2.5PN, 3PN, and 3.5PN
orbital phasing. We find phasing agreement consistent with internal
error estimates based on either approach, suggesting that PN waveforms
for this system are effective until the last orbit prior to final
merger.
\end{abstract}

\pacs{
04.25.Dm, 
04.25.Nx  
04.30.-w 
04.30.Db, 
95.30.Sf, 
97.60.Lf  
}

\maketitle

Compact astrophysical binaries spiral together due to the emission of
gravitational radiation.  Calculating the dynamics of these systems,
and the corresponding gravitational waveforms, has been a central
problem in general relativity for several decades.  With
first-generation interferometers such as LIGO, VIRGO,and GEO600 now
operating, and development moving forward on the space-based LISA
mission, accurate and reliable waveforms are urgently needed for
gravitational-wave data analysis.

Post-Newtonian (PN) methods, based on expansions in the parameter
$\epsilon \sim v/c$, have been the major analytic tool used to
calculate the system dynamics and waveforms during the early part of
the inspiral, when the binary components are relatively widely
separated and thus have a small orbital frequency \cite{Blanchet06}.
Currently, gravitational-wave data analysis for binary inspiral relies
on waveforms derived from PN methods \cite{BCV}.  The current
predicted orbital phase is available up to $O(\epsilon^{7})$, which is
referred to as 3.5PN order.  However the convergence properties of the
PN sequence are not well understood, and it is not yet clear how well
PN predictions work late in the inspiral when frequencies are high.

Numerical relativity, in which the full set of Einstein's equations is
solved on a computer, is needed to handle the final stages of the
binary evolution, when the components inspiral rapidly and merge.
Recently, there has been dramatic progress in the use of numerical
relativity to simulate the final inspiral and merger of black holes
\cite{Pretorius:2005gq,Campanelli:2005dd,Baker:2005vv,Campanelli:2006gf,
Baker:2006yw,Campanelli:2006uy,Campanelli:2006fg,Gonzalez:2006md,
Buonanno:2006ui}. These breakthroughs have allowed numerical
simulations with increasingly wider initial separations, producing
longer wavetrains.  Linking such simulations with the PN calculations
and comparing their waveforms in the late inspiral regime is a
pressing concern of gravitational-wave data analysis.  While
qualitative comparisons have suggested that the two methods agree
fairly well until shortly before the merger
\cite{Baker:2006yw,Buonanno:2006ui}, our group has found that even
inaccurate waveforms may approximately agree over several cycles.
Quantifying the level of agreement unambiguously requires
long, accurate simulations with very low eccentricity, and careful analysis.

We have carried out suitable numerical simulations of a merging
equal-mass, nonspinning black-hole binary. The black holes start on
nearly circular orbiting trajectories $\sim 1200M$ before merger,
where $M$ refers to the mass that the system would have had when the
black holes were still far apart, before radiative losses were
significant. $M$ is related to time by $M\equiv
5\times10^{-6}s(M/M_{\odot})$.  In this {\em Letter}, we
quantitatively compare crucial phasing information in our numerical
simulation waveforms with phasing in PN waveforms, finding striking
agreement.

The numerical simulations were performed using the moving puncture
method \cite{Campanelli:2005dd,Baker:2005vv,vanMeter:2006vi}.  We use
fourth-order Runge-Kutta time integration, fourth-order-accurate
finite spatial differencing, and second-order-accurate initial data.
Adaptive mesh refinement is used to resolve both the dynamics near the
black holes and the propagation of the gravitational waves
\cite{Baker:2006yw}.  The wave extraction sphere was of radius 60M and
the cubical outer boundary was of half-width 1536M; more details about
this simulation can be found in \cite{Baker:2006kr}.  We performed
physically equivalent runs at three different maximum resolutions: low
($3M/64$), medium ($3M/80$), and high ($M/32$).  We find fourth-order
convergence of the Hamiltonian constraint, and better than
second-order convergence of the momentum constraints during the runs.
 
The simulations begin at an angular gravitational-wave frequency
$\omega \sim 0.051M^{-1}$. The frequency then sweeps upward through
roughly an order of magnitude while the black holes undergo $\sim 7$
orbits, producing $\sim 14$ gravitational wave cycles before merger.
For such long-lasting simulations, the primary consideration in
providing a realistic initial data model is to set up the orbiting
black holes with minimal eccentricity, as gravitating binary systems
of comparable-mass objects are expected to circularize rapidly through
the emission of gravitational radiation.  We have selected an initial
black hole configuration with the relatively low eccentricity of less
than one percent, as measured below.
 
\begin{figure}
\includegraphics[scale=.28, angle=0]{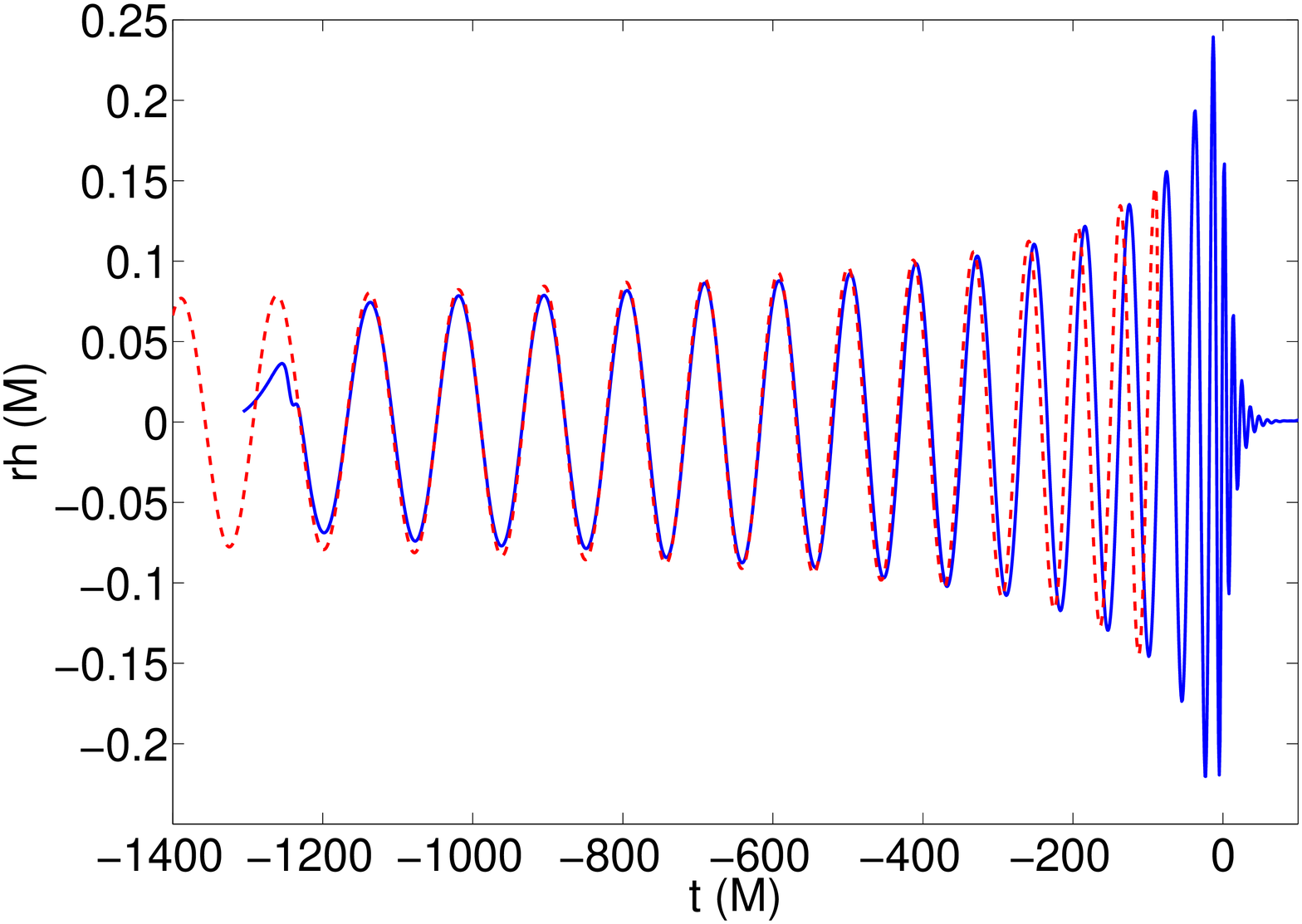}
\caption{Gravitational strain waveforms from the merger
of equal-mass Schwarzschild black holes. The solid curve is the
waveform from the high resolution numerical simulation, and the dashed
curve is a PN waveform with 3.5PN order phasing
\cite{Blanchet:2001ax,Blanchet:2004ek} and 2.5PN order amplitude
accuracy \cite{Arun:2004ff}. Time $t = 0$ is the moment of peak
radiation amplitude in the simulation.}
\label{fig:Wave}
\end{figure}

Fig.~\ref{fig:Wave} shows the gravitational wave strain generated by
our highest-resolution numerical run and that predicted by the PN
approximation with 3.5PN phasing
\cite{Blanchet:2001ax,Blanchet:2004ek} and 2.5PN (beyond leading
order) amplitude accuracy \cite{Arun:2004ff}. The waves are based on
the dominant $l=2,m=2$ spin-weighted spherical harmonic of the
radiation, and represent an observation made on the system's
equatorial plane, where only one polarization component contributes to
the measured strain.  The initial phase and time of the waves have
been adjusted so that the frequency and phase for each waveform agree
at a point, $t=-1000M$, that is early in the simulation, but after
transient effects from the initial data have subsided.  We will
quantify the phase agreement below using the frequency domain, so that
the time shifting, done for illustrative purposes in
Fig.~\ref{fig:Wave}, will have no impact on the subsequent analysis.

To conduct comparisons with PN calculations, we need to extract an
instantaneous gauge-invariant polarization phase $\phi$ and angular
frequency $\omega$ from our simulations.  These are derived from the
gravitational wave strain's first time-derivative, which is a robust
quantity in the numerical data.  This frequency corresponds to the
sweep rate of the polarization angle of the circularly polarized
gravitational wave that can be observed on the system's rotation axis.

We define eccentricity as a deviation from an underlying smooth, secular
trend.  We obtain a monotonic ``secular'' frequency-time relation by
modeling the waveform angular frequency $\omega$ as a fourth-order monotonic
polynomial $\omega_c(t)$, plus an eccentric modulation of the form
$d\omega(t) = \omega(t) -
\omega_c(t) = A \sin( \Phi(t))$, where $\Phi(t)$ is a quadratic
function of time.  Fitting this equation to our data yields
$A=8(\pm1)\times10^{-4} M^{-1}$.  For Keplerian systems, conserved
angular momentum is proportional to $r^2 \omega$, so the eccentricity
corresponds to half the fractional amplitude of the frequency
modulation: $e=A/(2\omega)$.  In our case the eccentricity starts near
$0.008$, decreasing by a factor of three by the time $\omega_c M \sim
0.15$.  We will compare our simulation with non-eccentric PN
calculations, with the expectation that small eccentricities have
minimal effect on the important underlying secular trend in the rate
at which frequency sweeps up approaching merger.

The phasing of the waveform is critical for gravitational wave
observation.  For data analysis, the optimal methods for both
detection and parameter estimation rely on matched filtering, which
employs a weighted inner product that can be expressed in Fourier
space as $<h,s>=\int df
[\tilde{h}^*(f)\tilde{s}(f)+\tilde{h}(f)\tilde{s}^*(f)]/S_n(f)$, where
${h}$ is the template being used, ${s}$ is the signal being analyzed,
and $S_n$ is the one-sided power spectral density of the detector's
noise \cite{Cutler93}.  A template that maximizes $<h,s>$ will provide
an optimal filter. Therefore, the most crucial factor is the relative
phasing of the template and signal.  The inner product will cease to
accumulate in sweeping through frequency if the template and the
signal evolve to be out of phase with each other by more than a
half-cycle, decreasing the effectiveness of the procedure.

Our key objective is to compare phasing between numerical and PN
waveforms.  We can make a stronger connection to the underlying
physics while avoiding issues with time alignment by comparing phases
as a function of polarization frequency, which corresponds to twice
the orbital frequency in the PN case.  For circular inspiral this
frequency should grow monotonically in time, with the frequency
$\omega_c$ providing a physical reference of the ``hardness'' of the
tightening binary.

Circular inspiral phasing information is typically derived in PN
theory by imposing an energy balance relation to deduce the rate at
which $\omega_c$ evolves from the radiation rate at a specified value
of $\omega_c$ \cite{Blanchet06}.  Though not strictly derived in the
PN context, this physically sensible condition currently allows the
determination of the \emph{chirp rate} $\dot\omega_c(\omega_c)$ up to
3.5PN order \cite{Buonanno:2006ui}.  From such a relation, information
about phase and time are determined by integrating
$d\phi/d\omega_c=\omega_c/\dot\omega_c$ and
$dt/d\omega_c=1/\dot\omega_c$.  The phasing information can be
represented by any one of several relations between phase, frequency
and time.  Various approaches take the PN-expanded representation of
one of these relations as the PN ``result'' for waveform phasing
\cite{Blanchet06,Buonanno:2006ui,Blanchet:2006gy}.  It has been
demonstrated \cite{Buonanno:2006ui} that the PN expansion of
$\dot\omega_c(\omega_c)$, numerically integrated as needed, has the
greatest utility for conducting comparisons of phasing with numerical
results during the late inspiral, and we adopt that convention.

For the purpose of comparison with our numerical simulations, we
invert the monotonic function $\omega_c(t)$ to obtain the phase as a
function of frequency: $\phi(\omega_c)=\phi(t(\omega_c))$. Note that
the effect of eccentricity is not removed from $\phi$, though the
``circularized'' frequency $\omega_c$ does provide the abscissa
according to which phases are compared in the different treatments.

\begin{figure}
\includegraphics[scale=.36, angle=-90]{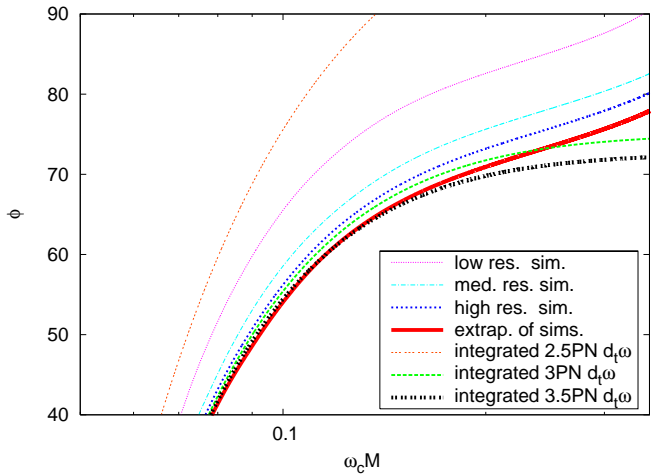}
\caption{Gravitational wave phase, in radians, for numerical and PN waveforms.
The solid curve is a Richardson extrapolation of the numerical
results.  The solid curve agrees well with the phase obtained by
numerically integrating the 3.5PN expansion of the chirp rate $\dot
\omega_c(\omega_c)$.  Each successive PN order shown agrees better
with the Richardson-extrapolated result, although this is not true of
all the preceding terms in the PN sequence, since the sequence does
not converge monotonically. }
\label{fig:phase-Richardsons}
\end{figure}

\begin{figure}
\includegraphics[scale=.32, angle=0]{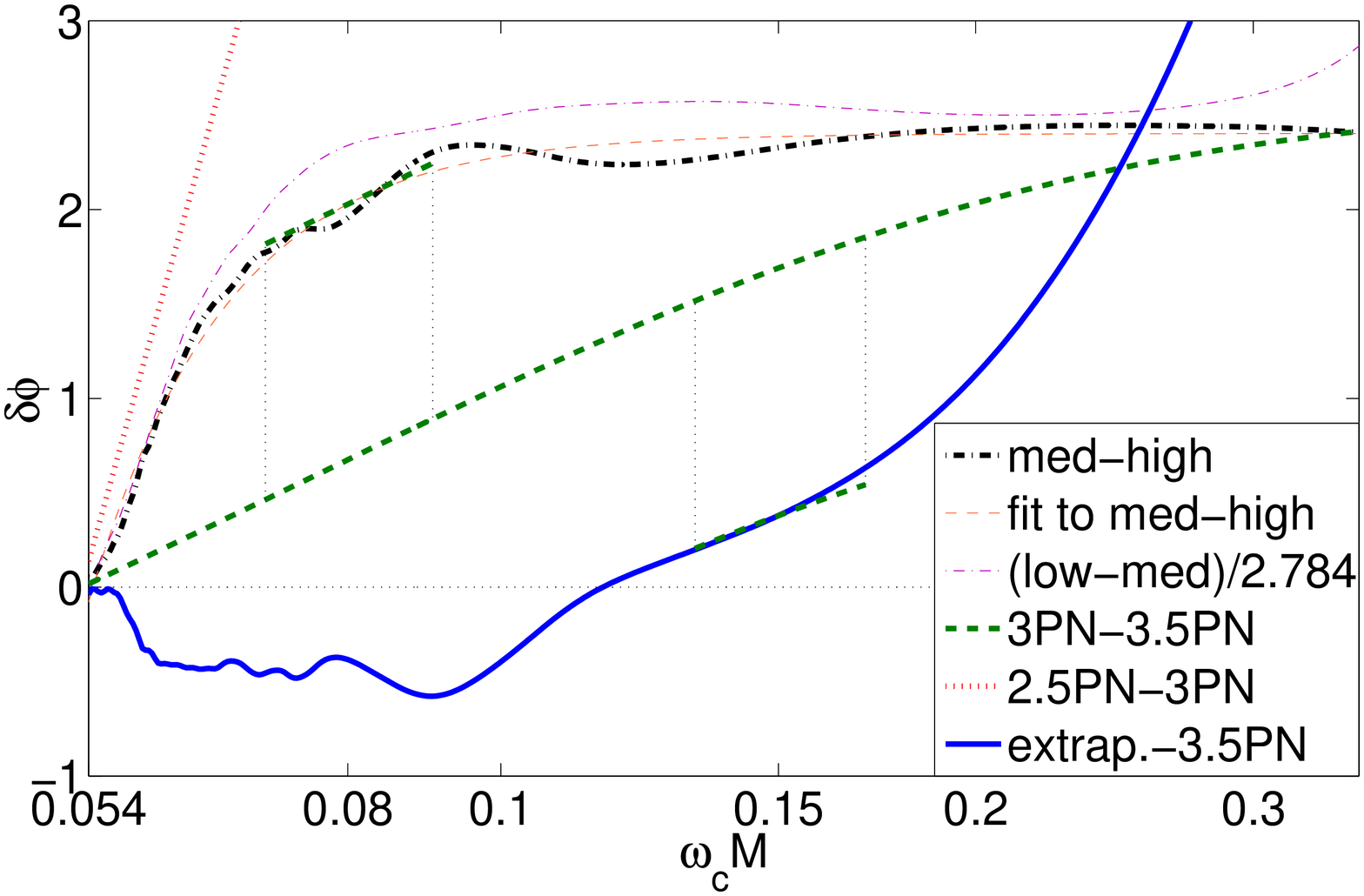}
\caption{Gravitational wave phase error estimates.
Differences between phasing from the integrated 3.5PN chirp rate and
Richardson extrapolation from the numerical simulations (solid curve)
are small, and are consistent with internal error estimates for the
numerical simulation results and the PN sequence.  Curves which
involve numerical phases are smoothed to remove high frequency noise.
}
\label{fig:phase-diff}
\end{figure}

Fig.~\ref{fig:phase-Richardsons} shows the wave phases as a function
of $\omega_c M$; here each phase is adjusted by addition of a constant
so that it vanishes at $\omega_c M=0.054$, corresponding to the time
$1000M$ before the radiation merger peak in the high-resolution
numerical simulation. As demonstrated in Fig. \ref{fig:phase-diff},
the sequence of numerical results converges at fourth order, allowing
us to obtain a fifth-order-accurate result by Richardson extrapolation
(thick solid line in Fig. \ref{fig:phase-Richardsons}).

In Fig.~\ref{fig:phase-Richardsons} we compare the numerical
simulation results for the phase with the numerically integrated PN
expansion of the chirp rate at 2.5PN, 3PN, and 3.5PN order.  The
agreement of the extrapolated numerical result with the integrated PN
chirp rate improves with each successive order, with the 3.5PN result
showing striking agreement up to about $\omega_c M \sim 0.15$.

We look more quantitatively at this agreement in
Fig.~\ref{fig:phase-diff} by plotting the phase differences $\delta
\phi$ that accumulate between different phase approximations, as a
function of frequency. The thick dash-dotted curve shows the phase
differences between our medium- and high-resolution results, while the
thin dash-dotted curve shows the differences between our low- and
medium-resolution results, scaled so that for fourth-order convergence
the curves should superpose.  This is indeed observed to good
approximation.  A good estimate for the error of the phase in the
high-resolution run is given by its difference from the phase obtained
by Richardson extrapolation; this comes out to $\sim 93 \%$ of the
med-high curve shown in the figure.  Note that the cumulative errors
in the numerically-generated waveforms accrue primarily at lower
frequencies, scaling approximately as $\omega_c^{-5}$; the thin dashed
curve shows a fit to the med-high curve with this scaling. The
deviations of the med-high curve from this fit show the effect of
eccentricity in our simulations. This accumulation of phase errors at
lower frequencies makes sense generally since the simulations spend
longer in that regime.
 
Without monotonic convergence between the 2PN, 2.5PN and 3PN at the
frequencies considered here, it is difficult to estimate errors in the
PN phase. Nonetheless we tentatively take the difference between the
integrated 3PN and 3.5PN chirp rates, shown by the dashed line in
Fig. \ref{fig:phase-diff}, as an upper bound on errors in a 3.5PN
waveform. We also show (thin dotted line) the accumulated phase
differences between the integrated 2.5PN and 3PN chirp rates to show
the relative contribution of the preceding PN term.

The trend in the slope of these error curves indicates the rate at
which phase error accumulates, as independently estimated within each
approach. The slope of the fit to the med-high error curve in
Fig. \ref{fig:phase-diff} is initially higher than that of the
3PN-3.5PN error curve, but decreases steadily, matching the PN
error-curve slope around $\omega_c M \sim 0.08$ and less
thereafter. For clarity a piece of the 3PN-3.5PN curve has been
translated upward to fit to the med-high curve in
Fig.~\ref{fig:phase-diff}.  This suggests that phasing errors for our
high-resolution simulation accumulate more quickly than 3.5 PN phasing
errors for $\omega_c M \lesssim 0.08$ ($t \lesssim -300M$), with the
numerical simulation phasing being more accurate than PN at higher
frequencies.  In both cases, the phase error accumulates to roughly
two radians by $\omega_c M \sim 0.15$ as the black holes begin to
plunge together.

We now address the central objective of this {\em Letter}, a
quantitative comparison of numerical and PN phasing results.  We
compare the Richardson-extrapolated phase, our best simulation
estimate, with the integrated 3.5PN chirp rate result.  The difference
is shown in the solid curve of Fig. \ref{fig:phase-diff}.  Note that,
overall, the phase differences between the PN and numerical results
(solid curve) accumulate less rapidly than the estimated errors in
either the high resolution simulation or the 3.5PN results over much
of the frequency range shown.  This illustrates that the numerical and
post-Newtonian predictions appear to be converging to a common answer
for waveform phase, in a frequency regime where the validity of the PN
predictions could not previously be assessed.  In the range $0.1
\lesssim\omega_c \lesssim 0.15$ the slope of the trend in the solid
curve surpasses that of the 3PN-3.5PN curve, with the precise
crossover point obscured by the effect of eccentric oscillations in
the numerical simulations; we have translated a portion of the
3PN-3.5PN curve down to the solid line for clarity.  After this range,
corresponding to $-170 \lesssim t/M \lesssim -70$, and occurring 1 to
3 wave cycles before the estimated end of the binary's last orbit, the
phase difference between the best estimates of the numerical and PN
approaches grows more significantly.

Through the frequency range $0.054 \lesssim \omega_c M \lesssim 0.15$
the net phase difference, measured against frequency, amounts to less
than one radian, a level of error that would be tolerable in many
gravitational wave data analysis applications.  Superficially, phase
differences in Fig.~\ref{fig:Wave} may appear smaller than the
differences quantified in Fig.~\ref{fig:phase-diff}.  This is because
the differences in chirp-rate are more directly evident in the
frequency-based phase comparisons.  This more immediate connection to
the merger dynamics, together with the avoidance of time-alignment
issues, makes frequency-based phase comparisons a more reliable
indicator of phasing differences.

Our results provide a crucial cross-validation of PN waveforms from
the late inspiral of binary merger, with results of new long-lasting
numerical simulations.  These simulations have sufficient accuracy to
provide a meaningful comparison with PN waveforms over the last $t\sim
1000M$ of the coalescence, specifically addressing a binary system of
equal-mass non-spinning black holes. We find phase agreement
consistent with internal phase-error estimates conducted in each
approach, indicating that phase accuracies within a few radians are
now achievable for this part of the coalescence waveform.

We emphasize, however, that there is still much important work to be
done in improving and further assessing PN and numerical simulation
waveforms.  Certainly we have only addressed one case in a large
parameter space of potential binaries, which will inevitably include
systems, such as rapidly precessing unequal-mass spinning binaries,
that are harder to treat with present PN and numerical techniques.
With either approach, even for our simple case, a non-negligible
amount of phasing error accumulates over the range studied, and more
will have accumulated at lower frequencies addressable through the PN
approximation. We expect continuing developments in numerical
simulations and the pursuit of higher-order PN treatments to be
crucial for developing a refined understanding of coalescence
waveforms, which will be crucial in some data analysis applications
for gravitational wave observations.

\acknowledgments

This work was supported in part by NASA grant O5-BEFS-05-0044.  The
simulations were carried out using Project Columbia at NASA Ames
Research Center.  J.v.M. and B. K. were supported by the NASA
Postdoctoral Program at the Oak Ridge Associated Universities.

%

\end{document}